\documentclass[twocolumn,superscriptaddress,amssymb,amsmath,nobibnotes,aps,prd,showpacs,nofootinbib]{revtex4}%
\pdfoutput=1

\usepackage{graphicx}
\usepackage{epsf}
\usepackage{bm}
\usepackage{amsmath,hyperref}
\usepackage{amsfonts}
\usepackage{amssymb}
\usepackage{epstopdf}
\usepackage{natbib}%
\usepackage{rotating}
\usepackage{pdflscape}
\usepackage{adjustbox}
\usepackage[toc,page]{appendix}
\providecommand{\U}[1]{\protect\rule{.1in}{.1in}}
\newcommand{\be}{\begin{equation}}
\newcommand{\ee}{\end{equation}}

\newcommand{\mincir}{\raise
-3.truept\hbox{\rlap{\hbox{$\sim$}}\raise4.truept\hbox{$<$}\ }}
\newcommand{\magcir}{\raise
-3.truept\hbox{\rlap{\hbox{$\sim$}}\raise4.truept\hbox{$>$}\ }}

\usepackage{color}

\begin{document}

\title{Probing the properties of relic neutrinos using the cosmic microwave
background, the \textit{Hubble Space Telescope} and galaxy clusters}

\author{Rafael C. Nunes}
\email{rafadcnunes@gmail}
\affiliation{Departamento de F\'isica, Universidade Federal de Juiz de Fora, 36036-330,
Juiz de Fora, MG, Brazil}

\author{Alexander Bonilla}
\email{abonilla@fisica.ufjf.br}
\affiliation{Departamento de F\'isica, Universidade Federal de Juiz de Fora, 36036-330,
Juiz de Fora, MG, Brazil}

\pacs{98.80.-k; 98.80.Es}

\begin{abstract}

We investigate the observational constraints on the cosmic neutrino background (CNB) given by the extended $\Lambda$CDM scenario
($\Lambda$CDM $+ N_{\rm eff} + \sum m_{\nu} + c^2_{\rm  eff} + c^2_{\rm  vis} + \xi_{\nu}$)  using the latest observational data from \textit{Planck} CMB (temperature power spectrum, low-polarisation and lensing reconstruction), baryon acoustic oscillations (BAOs), the new recent local value of the Hubble constant from\textit{ Hubble Space Telescope} (\textit{HST}) and information of the abundance of galaxy clusters (GCs). We study the constraints on the CNB background using CMB + BAO + \textit{HST} data with and without the GC data. We find $\Delta N_{\rm eff} = 0.614 \pm 0.26$ at 68 per cent confidence level when the GC data are added in the analysis. We do not find significant deviation for sound speed in the CNB rest frame. We also analyze the particular case $\Lambda$CDM $+ N_{\rm eff} + \sum m_{\nu} + \xi_{\nu}$ with the observational data. Within this scenario, we find $\Delta N_{\rm eff} = 0.60 \pm 0.28$ at 68 per cent confidence level. In both the scenarios, no mean deviations are found for the degeneracy parameter.

\end{abstract}

\maketitle

\section{Introduction}

The existence of a cosmic neutrino background (CNB), which comprises the so-called relic neutrinos, 
is a consequence of the thermal history of the universe where the neutrinos decouple from the rest of the cosmic plasma at
$k_B T \sim$ MeV and start streaming freely. Unlike the cosmic microwave background (CMB), the CNB is yet to be detected directly,
and such a direct detection proves to be difficult \cite{CNB1}. However, indirect measures
have been established by using CMB as well as estimations from the primordial abundances of
light elements. Recently,  Follin et al. \cite{CNB2} have interpreted data about damping of acoustic oscillations of the CMB
and they have demonstrated a detection of the temporal phase shift generated by neutrino perturbations. 
This detection is the most model-independent determination of the existence of CNB. 

The properties the massive neutrinos play an important role in the dynamics of the universe,
inferring direct changes in important cosmological sources and, consequently, in 
the determination of cosmological parameters; see \cite{Dolgov,Lesgourgues,Abazajian} for reviews. 
The effects of the relic neutrinos on the CMB and LSS are only gravitational, as they are decoupled (free-streaming particles) 
at the time of recombination and structure formation. The standard parameters that characterize these effects on
cosmological sources are the effective number of species $N_{\rm eff}$ and the total neutrino mass $\sum m_{\nu}$.
\textit{Planck} team \cite{Planck2015} within the $\Lambda$CDM + $\sum m_{\nu}$ model has constrained 
$\sum m_{\nu} < 0.194$ eV (from the CMB alone), and $N_{\rm eff}= 3.04 \pm 0.33$ at 95 per cent confidence level (CL).
The value of $N_{\rm eff}$ via theoretical calculations is well determined within the framework of the 
standard model, namely  $N_{\rm eff}=3.046$ \cite{book}. The evidence of any positive deviation from this value can be a signal
that the radiation content of the Universe is not only due to photons and neutrinos, 
but also to some extra relativistic relics, the so-called dark radiation in the literature, and parametrized 
by $\Delta N_{\rm eff} = N_{\rm eff}-3.046$.

However, two phenomenological parameters $c^2_{\rm eff}$ and $c^2_{\rm vis}$ are also introduced
to infer properties of the CNB. Here, $c^2_{\rm eff}$ is the sound speed in the CNB rest frame 
and $c^2_{\rm vis}$ is the viscosity parameter, which parameterizes the anisotropic stress.
The evolution of standard neutrinos (non-interacting free-streaming neutrinos) is obtained for $c^2_{\rm eff}=c^2_{\rm vis}=1/3$. 
The CNB properties, including constraints on $c^2_{\rm eff}$ and $c^2_{\rm vis}$, 
have been investigated via different methods and approaches \citep{CNB3,CNB4,CNB5,CNB6,CNB7,CNB8,CNB9,CNB10,Benjamin}.
It is important to mention that these parametrizations were strongly inspired by pioneer works 
about dark matter properties \cite{Hu1,Hu2}. From the temperature power spectrum (TT), the temperature-polarization cross spectrum (TE), the polarization power spectrum (EE) + low-polarization (lowP) + baryon acoustic oscillations (BAOs), the \textit{Planck} collaboration \cite{Planck2015} has constrained $c^2_{\rm eff} = 0.3242 \pm 0.0059$ and $c^2_{\rm vis} = 0.31 \pm 0.037$. 
Recently, within the $\Lambda$CDM $+ c^2_{\rm eff} + c^2_{\rm vis} + \sum m_{\nu}$ model,  \cite{Benjamin}
have reported the constraints $c^2_{\rm eff} = 0.309 \pm 0.013$ and $c^2_{\rm vis} = 0.54^{+0.17}_{-0.18}$ at 95 per cent CL from CMB + lensing + BAO data. In general terms, measuring a deviation from $(c^2_{\rm eff},c^2_{\rm vis})=1/3$ can refute the null hypothesis
that the relic neutrinos are relativistic and free-streaming.

Another natural extension of the physics properties of the neutrino is to consider 
a certain degree of lepton asymmetry (a cosmological leptonic asymmetry), which is usually parametrized by the so-called
degeneracy parameter $\xi_{\nu}=u_{\nu}/T_{\nu0}$ \citep{Dolgov,xi1,xi2,xi3,Dominik01}, where 
$u_{\nu}$ is the neutrino chemical potential and $T_{\nu0}$ is the current temperature of the CNB, $T_{\nu0}/T_{CMB} = (4/11)^{1/3}$.
The leptonic asymmetry also shifts the equilibrium between protons and neutrons at the Big Bang Nucleosynthesis (BBN) epoch, leading to indirect
effects on the CMB anisotropy through the primordial helium abundance $Y_{He}$. The effects of the massive neutrinos and a leptonic 
asymmetry on BBN and CMB have been investigated in many studies \cite{xi4,xi5,xi6,xi7,xi8,xi9,xi10,xi11,xi12,xi13,Dominik02}.
Recently, a leptonic asymmetric model of cosmological data are reported at 95 per cent CL by \cite{xi11}.
In the Appendix, we present a brief review/description on the neutrino cosmological leptonic asymmetry.

In this paper, we consider the extension of the minimal $\Lambda$CDM model to $\Lambda$CDM 
$+ N_{\rm eff} + \sum m_{\nu} + c^2_{\rm  eff} + c^2_{\rm  vis} + \xi_{\nu}$ and we derive observational constraints 
on the additional five neutrinos parameters that can characterize the properties of the CNB. For this purpose,
we consider data from CMB observed by \textit{Planck} 2015, BAOs, the recent local value of the Hubble constant from \textit{Hubble Space Telescope} (\textit{HST}) and information from the abundance of galaxy clusters (GCs).

The paper is organized as follows. In Section \ref{CNB}, we briefly comment on the CNB. In Section \ref{model}, we present
the models and the data sets used in this work. In Sections \ref{results} and \ref{conclusions}, we present our results and conclusions, 
respectively. 



\section{Cosmic neutrino background}
\label{CNB}

The total radiation density energy can be parametrized (when the neutrinos are relativistic) by

\begin{eqnarray}
\rho_{r} = \Big[1 + \frac{7}{8}\Big(\frac{4}{11} \Big)^{4/3} N_{\rm eff} \Big] \rho_{\gamma},
\end{eqnarray}

where the factor 7/8 appears because neutrinos are fermions. Neutrinos become non-relativistic when their average momentum falls below
their mass. In the Appendix, we present a brief description of the massive degenerate neutrinos.

Some approximations for massive neutrino have been discussed
in the literature \cite{Hu1,Ma_Bertschinger,Lewis,Komatsu}. Here, let us follow the methodology 
and notation of \cite{Lesgourgues2}, where an extension of the ultrarelativistic 
fluid approximation is presented. Within the fluid approximation,
the continuity, Euler, and shear equations, in the synchronous gauge, are given by

\begin{eqnarray}
\label{nu1}
\dot{\delta}_{\nu} = -(1+w_{\nu})\Big( \theta_{\nu} + \frac{\dot{h}}{2} \Big) - 3\frac{\dot{a}}{a}
(c^2_{\rm eff}-w_{\nu})\delta_{\nu} + \nonumber \\
9 \Big(\frac{\dot{a}}{a} \Big)^2 (1+w_{\nu})(c^2_{\rm eff} - c^2_{g})\frac{\theta_{\nu}}{k^2},
\end{eqnarray}

\begin{eqnarray}
\label{nu2}
\dot{\theta}_{\nu} = -\frac{\dot{a}}{a}(1-3c^2_{\rm eff})\theta_{\nu} + 
\frac{c^2_{\rm eff}}{1+w_{\nu}}k^2 \delta - k^2 \sigma_{\nu},
\end{eqnarray}

and

\begin{eqnarray}
\label{nu3}
\dot{\sigma}_{\nu} = -3\Big(\frac{1}{\tau} + \frac{\dot{a}}{a}
\Big[\frac{2}{3} - c^2_g - \frac{p_{pseudo}}{3p}\Big] \Big) \sigma_{\nu} + \nonumber \\
\frac{8}{3}\frac{c^2_{\rm vis}}{1+w_{\nu}}\Big[ \theta_{\nu} + \dot{h} \Big].
\end{eqnarray}

In equations (\ref{nu1})-(\ref{nu3}), $w_{\nu} = p_{\nu}/\rho_{\nu}$ (which starts with $w_{\nu} = 1/3$ at early times and drops to $w_{\nu} \simeq 0$ 
when neutrinos become nonrelativistic), $c^2_{\rm vis} = 3 w_{\nu} c^2_g$ and

\begin{eqnarray}
c^2_g = \frac{w_{\nu}}{3+3w_{\nu}}\Big( 5 - \frac{p_{pseudo}}{p} \Big),
\end{eqnarray}
where the quantity $p_{pseudo}$ is the so-called the pseudo-pressure. See \cite{Lesgourgues2} and \cite{Benjamin} for details of the Boltzmann hierarchy.

In the application of the eqs. (\ref{nu1})-(\ref{nu3}), we consider three active neutrinos: one massive neutrino $\nu_1$ and two massless neutrinos $\nu_2$ and $\nu_3$, which is standard practice in the literature.
Because here we have the standard $\Lambda$CDM scenario, the baryons, cold dark matter, and photons
follow the standard evolution (both at the background and perturbation levels).

\section{Models and data analysis}
\label{model}

We consider two different models. First, let us take $\Lambda$CDM + $N_{\rm eff} + \sum m_{\nu} + c^2_{\rm  eff} 
+ c^2_{\rm  vis} + \xi$ (\textbf{Model I}). Then, we take a particular case of the Model I 
when $c^2_{\rm  eff} = c^2_{\rm  vis} = 1/3$, i.e., $\Lambda$CDM + $N_{\rm eff} + \sum m_{\nu} + \xi$ (\textbf{Model II}).
Following the \textit{Planck} collaboration, we fix the mass ordering of the active neutrinos 
to the normal hierarchy with the minimum masses allowed by oscillation experiments, i.e., $\sum m_{\nu} = 0.06$ eV.
In this work, we consider one massive neutrino flavour $\nu_1$ and two massless flavours $\nu_2$, $\nu_3$  
with degeneracy parameter $\xi_{\nu} = \xi_{\nu1} = \xi_{\nu2} = \xi_{\nu3}$.
In order to constrain the free parameters of the models, we consider the following data sets:
\\

CMB: We consider a conservative data set from \textit{Planck} 2015 comprised of the likelihoods of temperature power spectrum (TT), 
low-polarisation and lensing reconstruction.

BAO: The BAO measurements from the  Six  Degree  Field  Galaxy  Survey  (6dF) \cite{bao1}, 
the  Main  Galaxy  Sample  of  Data  Release 7  of  Sloan  Digital  Sky  Survey  (SDSS-MGS) \cite{bao2}, 
the  LOWZ  and  CMASS  galaxy  samples  of  the Baryon  Oscillation  Spectroscopic  Survey  (BOSS-LOWZ  and  BOSS-CMASS,  
respectively) \cite{bao3},  and the distribution of the LymanForest in BOSS (BOSS-Ly) \cite{bao4}. 
These data points are summarized in table I of \cite{bao5}.

\textit{HST}: We include the new local value of Hubble constant, $H_0 = 73.02 \pm 1.79$ km/s/Mpc as measured by \cite{riess} 
with a 2.4 per cent determination.

GC: The measurements from the abundance of GCs are a powerful probe of the growth of cosmic structures.
However, this cosmological test depends on the calibration of the mass-observable relation, which can represents uncertainty 
in the measure of clusters samples properties. It is well known that cluster data are in tension with CMB data up to 95 per cent CL, 
especially when taking into account contributions due to non-linear scales. In orden to explore the full GC counts as 
a cosmological probe, it is necessary to take into account the modelling the number of haloes within a redshift and mass bin, for example. 
This modelling is hard and expensive to perform. However, the cosmological information enclosed in the cluster abundance is efficiently parametrized by $S_8 = \sigma_8 (\Omega_m/\alpha)^{\beta}$, where $\sigma_8$ is the linear amplitude of fluctuations on 8 Mpc/h scale and 
$\alpha$ and $\beta$ are the fiducial value adopted in each survey analysis. It can be an exhausting task to analyze different clusters samples in order to verify possible systematic effects that might exist in each survey. \cite{Ruiz} show that cluster abundance (from the full expression) carries less information about geometry parameters than about growth of structures (to constrain the growth parameters). This consideration justifies the choice of using CG data in the plan $S_8$, which also minimizes the computational cost. This methodology was also recently adopted by \cite{Bernal}. Table~\ref{tab1} summarizes the measures of $S_8$ used in this work. 
\\

\begin{table*}
\centering
 \caption{Cluster abundance measurements given in terms of $S_8$ included in our analysis.}
         \label{tab1}
          \begin{tabular}{lcccr}
          \hline
           Type            &  $\alpha$     &  $\beta$   &  Measurement  & Reference  \\
          \hline
          Number counts             & 1.0   &  0.5    & 0.465 $\pm$ 0.03   & \cite{CG1} \\
          Number counts             & 0.25  &  0.41   & 0.832 $\pm$ 0.03   & \cite{CG2} \\
          X-ray counts              & 0.32  &  0.30   & 0.86  $\pm$ 0.04   & \cite{CG3} \\
          Sunyaev-Zeldovich effect  & 0.25  &  0.298  & 0.785 $\pm$ 0.037  & \cite{CG4} \\
          X-ray cross CMB           & 0.30  &  0.26   & 0.80  $\pm$ 0.02   & \cite{CG5} \\
          X-ray luminosities        &  0.30 &  0.25   & 0.80  $\pm$ 0.04   & \cite{CG6} \\
          Sunyaev-Zeldovich effect  &  0.27 &  0.301  & 0.782 $\pm$ 0.01   & \cite{CG7}  \\  
          X-ray masses              &  0.25 &  0.47   & 0.813 $\pm$ 0.013  & \cite{CG8}  \\
          Tomographic weak lensing  &  0.27 &  0.46   & 0.774 $\pm$ 0.040  & \cite{CG9} \\  
          \hline
          \end{tabular}
\end{table*}

We use the publicly available CLASS \citep{class} and Monte Python \citep{monte} codes for constraining parameters of the models considered in the present work. We use Metropolis Hastings algorithm with uniform priors on the model parameters to obtain correlated Markov Chain Monte Carlo samples by considering two combinations of data sets: CMB + BAO + \textit{HST} and CMB + BAO + \textit{HST} + GC. All the parameter chains in our analysis converge according to the Gelman-Rubin criteria $1 - R < 0.01$ \citep{Gelman}.

\begin{table}
\centering
\caption{Constraints at 68 and 95 per cent CLs on some parameters of the Model I. The parameter $H_0$ is in the units of km s${}^{-1}$ Mpc${}^{-1}$ and $\sum m_{\nu}$ is in units of eV.}
\label{tab2}
\begin{tabular} { l l l}
\hline
  Parameter                     &  CMB + BAO + $H_0$ &  CMB + BAO + $H_0$ + GC  \\
\hline

{$\sum m_{\rm \nu}      $} & $<0.24 \, (< 0.36)  $  & $ <0.64 \, (< 0.81) $\\\\

{$c^2_{\rm vis}     $} & $0.63^{+0.17+0.32}_{-0.17-0.32}      $  & $0.58^{+0.22+0.40}_{-0.25-0.40}$   \\\\

{$c^2_{\rm eff}    $} & $0.311^{+0.012+0.028}_{-0.015-0.027}   $   &  $0.319^{+0.013+0.024}_{-0.013-0.027}   $\\\\

{$\xi_\nu    $} & $0.1^{+0.54+1.0}_{-0.54-1.0}         $             &  $0.02^{+0.50+0.90}_{-0.50-0.85}      $\\\\

{$N_{\rm eff}     $} & $3.41^{+ 0.23 \, + 0.43}_{-0.23 \, -0.42}      $    &  $3.66^{+0.26 \, +0.48}_{+0.26 \, -0.49}      $\\\\

{$\Omega_{\Lambda }$} & $0.706^{+0.008+0.016}_{-0.008-0.016}   $   & $0.706^{+0.008+0.015}_{-0.008-0.015}   $\\\\

{$Y_{He}            $} & $0.2523^{+0.0029+0.0054}_{-0.0029-0.0056}$  & $0.2557^{+0.0032+0.0059}_{-0.0032-0.0063}$\\\\

{$H_0             $} & $69.8^{1.3+2.5}_{1.3-2.5}        $     &  $70.7^{+1.2+2.4}_{-1.2-2.2}        $\\\\

{$\sigma_8        $} & $0.839^{+0.018+0.036}_{-0.018-0.037}   $    &  $0.776^{+0.010+0.019}_{-0.010-0.019}   $\\

\hline
\end{tabular}
\end{table}

\begin{table}
\centering
\caption{Constraints at 68 and 95 per cent CLs on some parameters of the Model II. The parameter $H_0$ is in the units of km s${}^{-1}$ Mpc${}^{-1}$ and $\sum m_{\nu}$ is in units of eV.}
\label{tab3}
\begin{tabular} { l l l}
\hline
  Parameter                     &  CMB + BAO + $H_0$ &  CMB + BAO + $H_0$ + GC  \\
\hline

{$\sum m_{\rm \nu}      $} & $< 0.18 \, (< 0.30) $ & $ < 0.52 \, (< 0.64)      $\\\\

{$\xi_\nu    $} & $0.05^{+0.56+0.97}_{-0.56-0.99}      $             & $-0.02^{+0.51+0.92}_{-0.51-0.89}     $\\\\

{$N_{\rm eff}           $} & $3.49^{+0.21 \, +0.44}_{-0.23 \, -0.42}      $ & $3.65^{+0.28 \, +0.57}_{-0.28 \, -0.60}      $\\\\

{$\Omega_{\rm \Lambda }$} & $0.703^{+0.009+0.015}_{-0.008-0.016}   $ &  $0.706^{+0.008+0.015}_{-0.008-0.016}   $\\\\

{$Y_{He}            $} & $0.2537^{+0.0028+0.0056}_{-0.0028-0.0056}$ &  $0.2557^{+0.0038+0.0071}_{-0.0032-0.0077}$\\\\

{$H_0             $} & $70.5^{+1.3+2.7}_{-1.3-2.6}        $ &  $71.2^{+1.4+2.6}_{-1.4-2.7}        $\\\\

{$\sigma_8        $} & $0.823^{+0.016+0.030}_{+0.014-0.032}   $ & $0.777^{+0.010+0.020}_{-0.010-0.019}   $\\

\hline
\end{tabular}
\end{table}

\begin{figure}
   	\includegraphics[width=9cm]{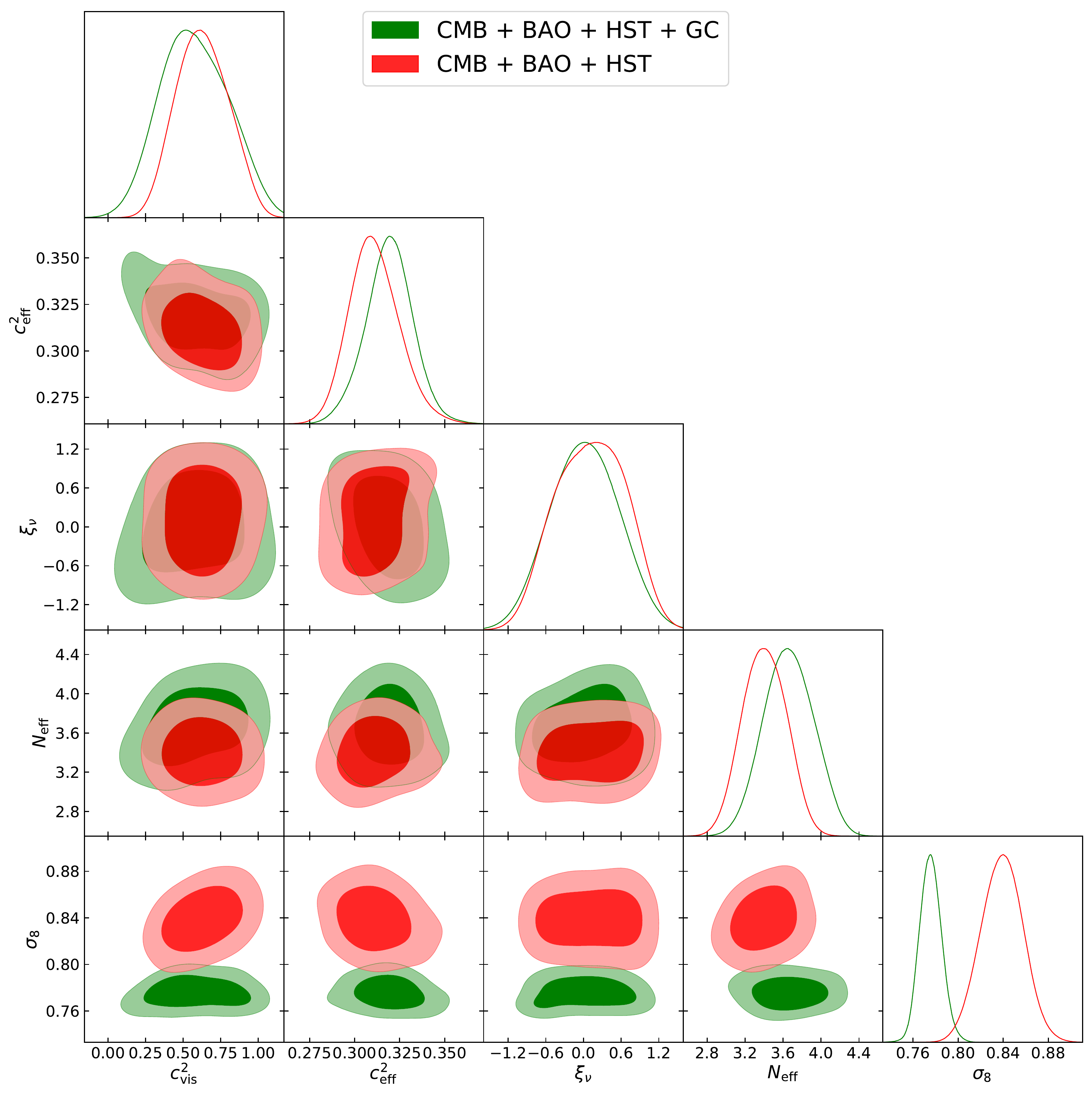}
   	\caption{One-dimensional marginalized distribution and 68 and 95 per cent CLs regions for some selected parameters of the 
   	         Model I. }
   	\label{Contours}
\end{figure}

\begin{figure}
   	\includegraphics[width=8cm]{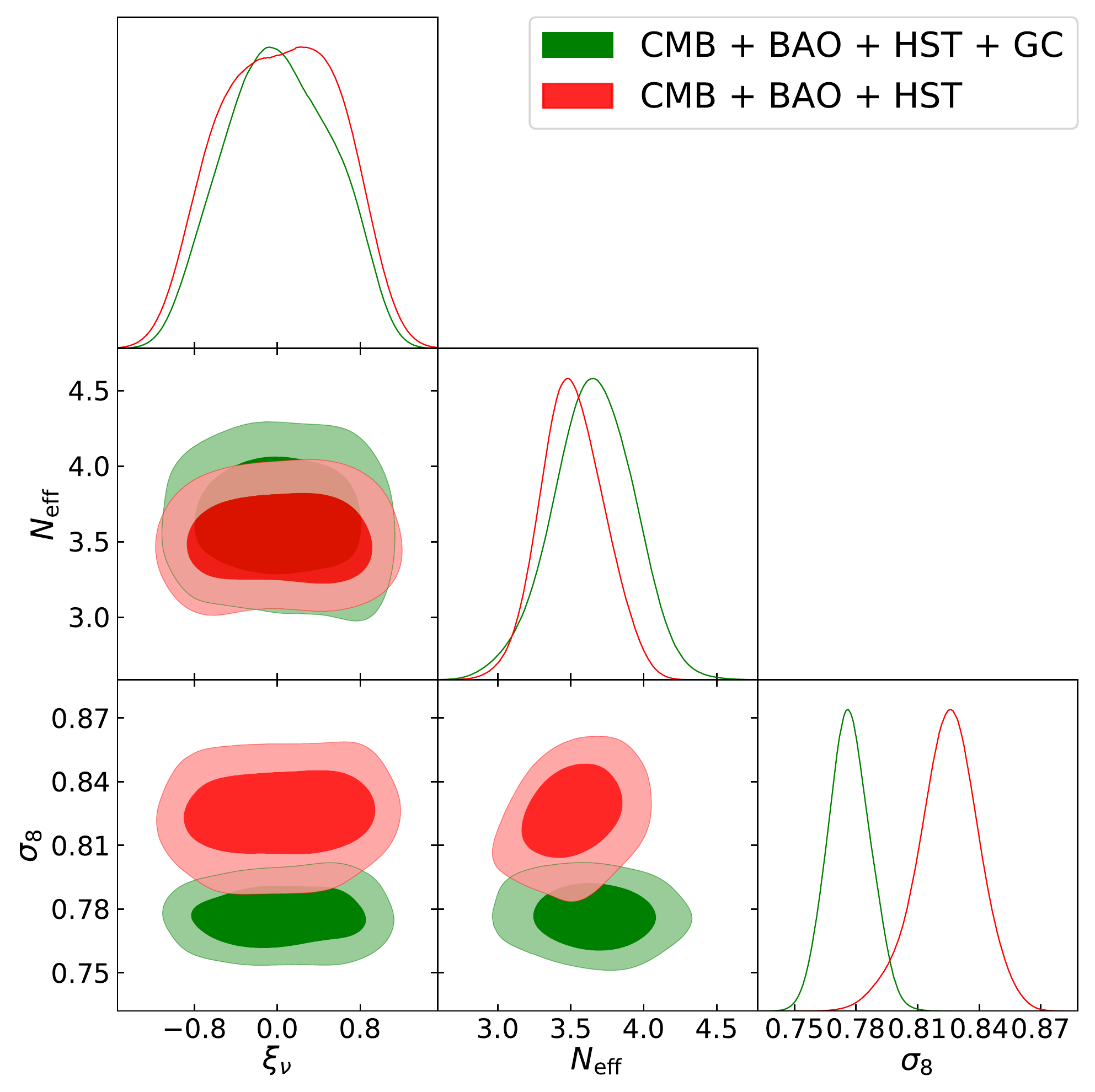}
   	\caption{One-dimensional marginalized distribution and 68 per cent CL and 95 per cent CL regions for some selected parameters of the 
   	         Model II. }
   	\label{Contours2}
\end{figure}

\begin{figure*}
   	\includegraphics[width=6cm,height=5cm]{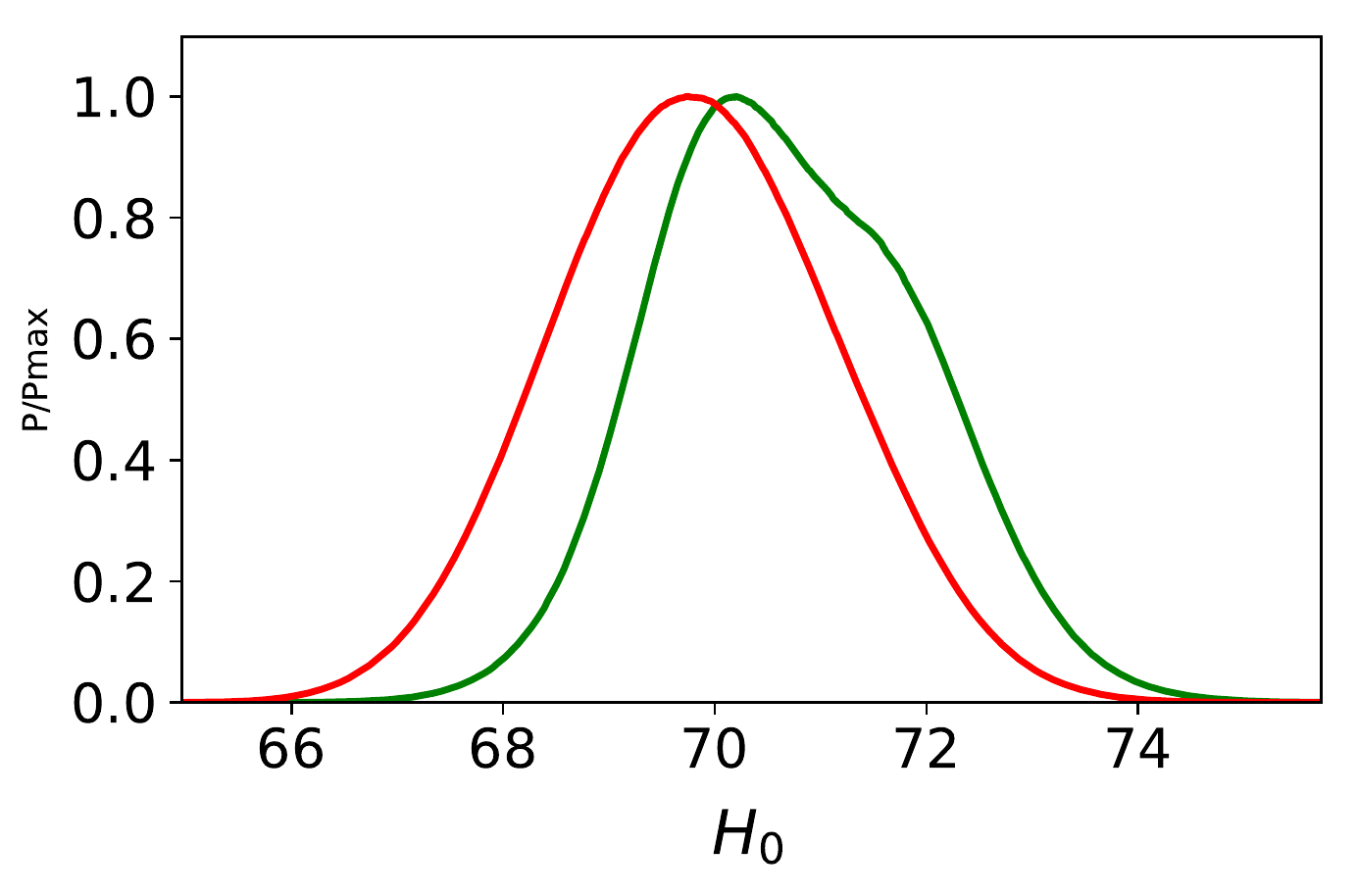}
   \includegraphics[width=6cm,height=5cm]{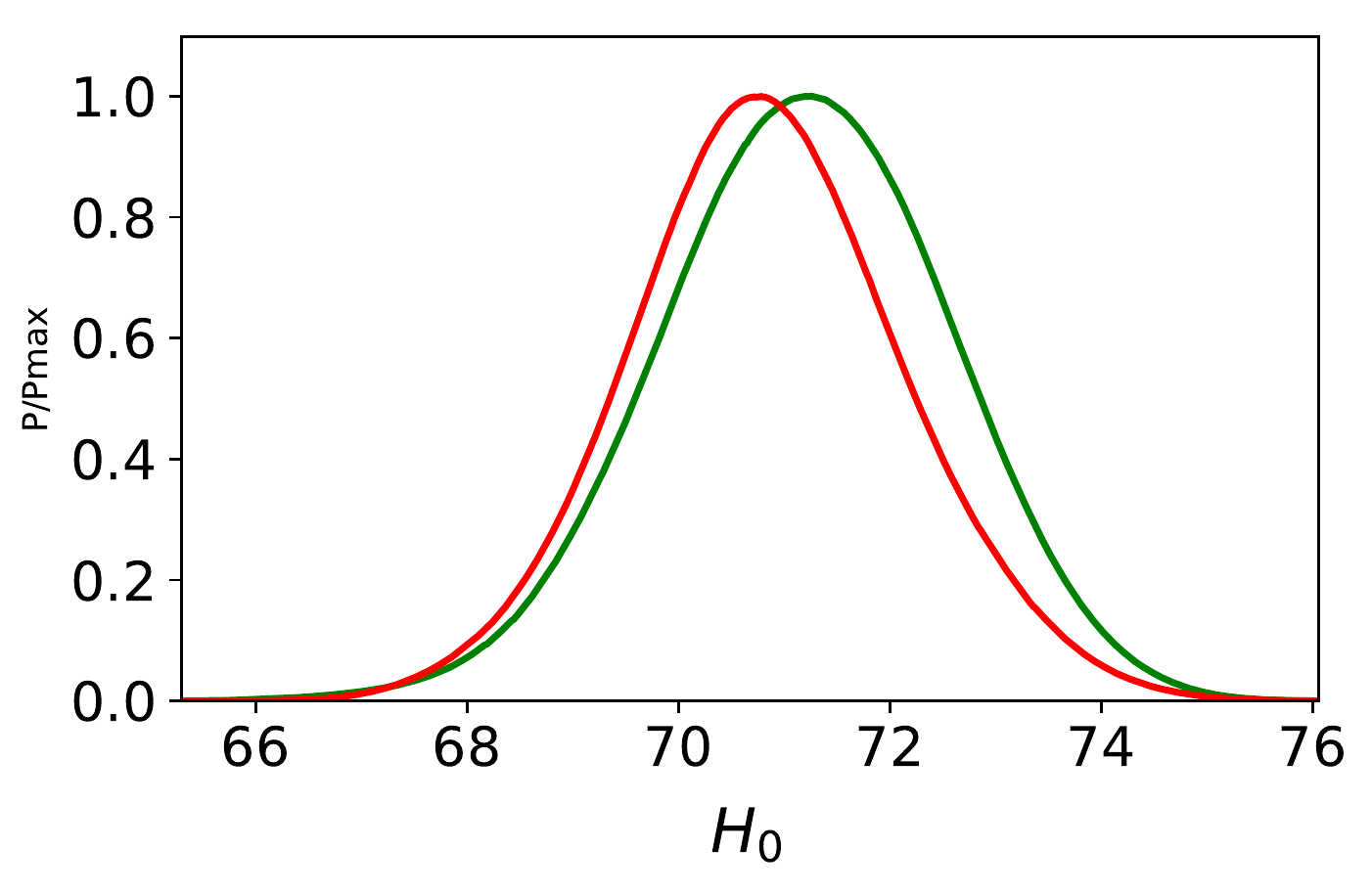}
   \caption{The likelihoods of the parameter $H_0$ for Model I (left panel) and Model II (right panel), in red (CMB + BAO + \textit{HST}) and 
   green (CMB + BAO + \textit{HST} + GC).}
   	\label{HS1}
\end{figure*}

\begin{figure}
    	\includegraphics[width=8cm,height=7cm]{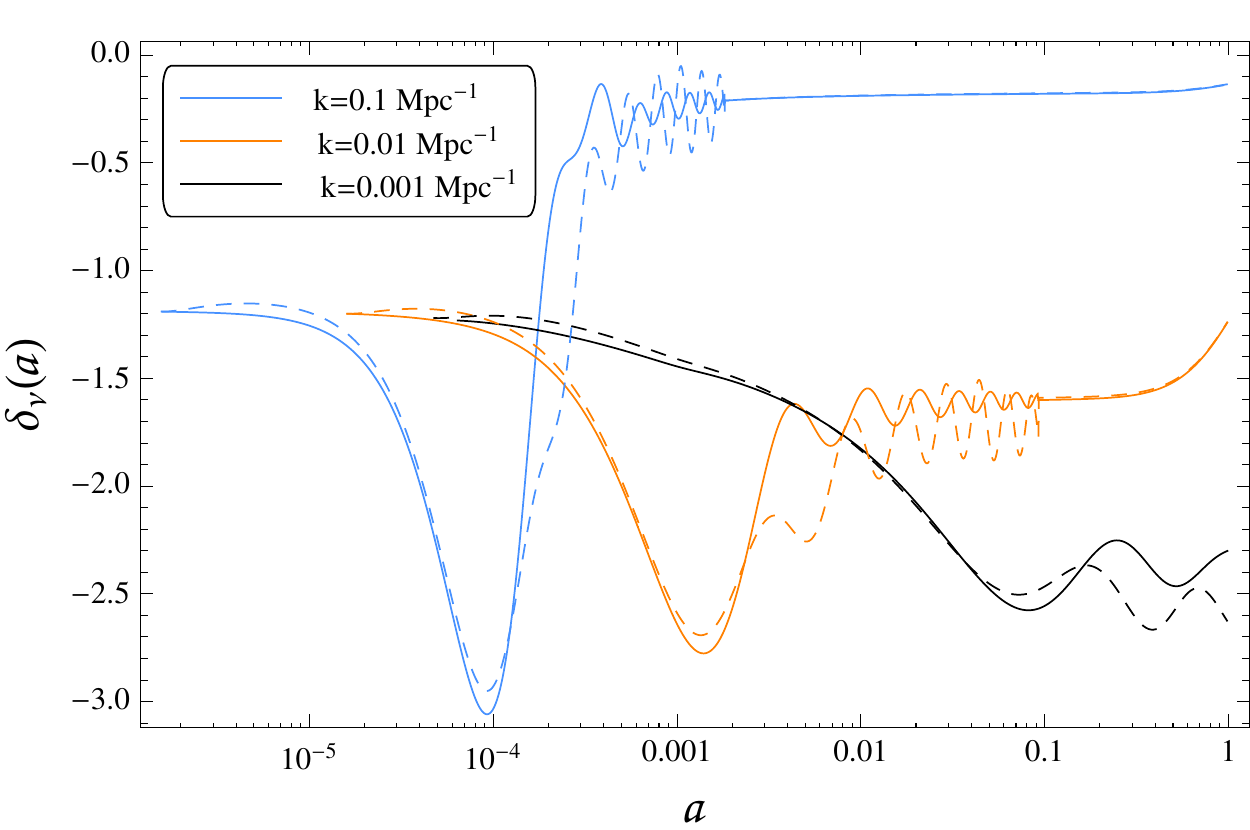}
    \caption{Neutrino density perturbations as a function of scale factor for three fixed scales,  
    	$k =$ 0.001 Mpc$^{-1}$, 0.01 Mpc$^{-1}$ and 0.1 Mpc$^{-1}$. In drawing the graphs we have taken the best fit values from 
    	our analysis. The continuous and dashed line represent the models I and II from CMB + BAO + \textit{HST} + GC, respectively.}
    	\label{delta1}
\end{figure}

\begin{figure}
     	\includegraphics[width=8cm,height=7cm]{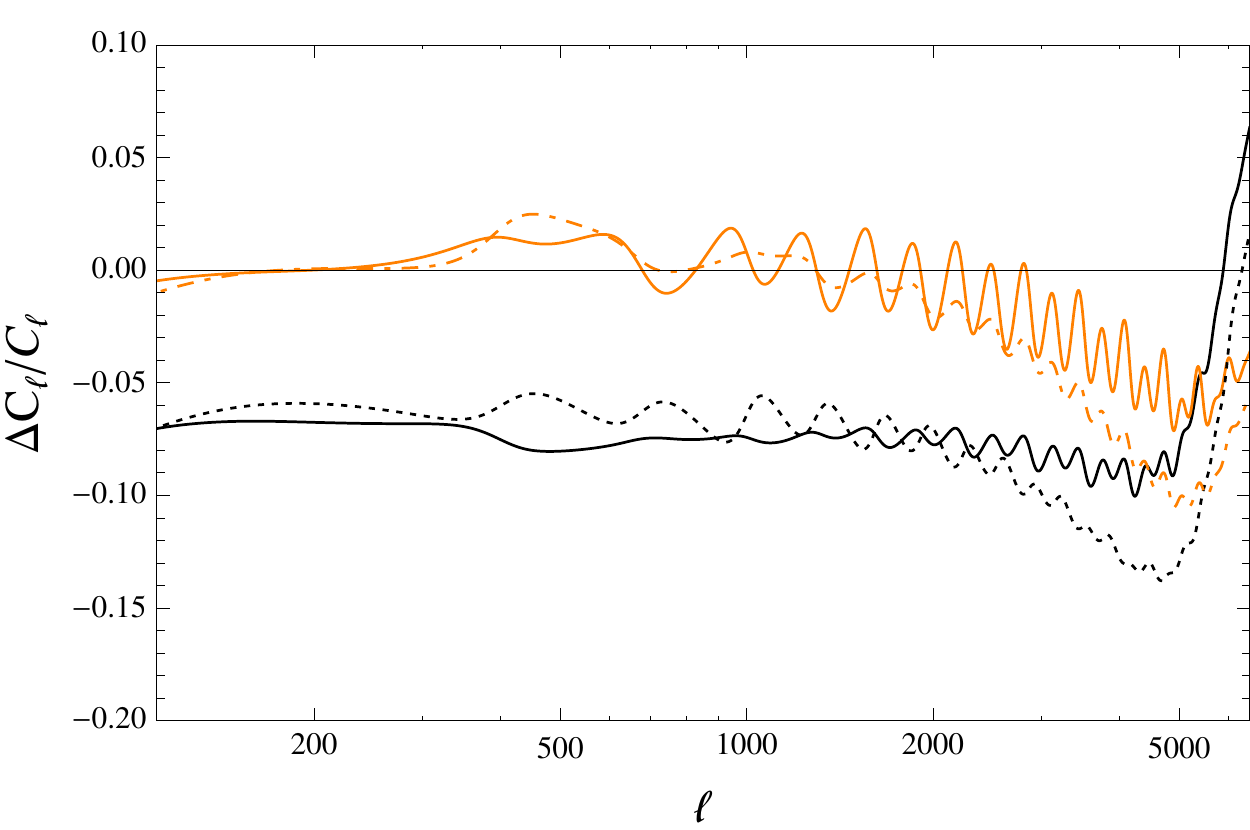}
    \caption{CMB temperature power spectrum difference among the models investigated here and the six parameter $\Lambda$CDM model, i.e.,
             $ \Delta C_l/C_l = (C^{\Lambda CDM \, \, extended}_l - C^{\Lambda CDM \, \,  Planck \, \, 2015}_l)/C^{\Lambda CDM \, \, Planck \, \,  2015}_l$.
             The black continuous and dotted lines represent the Model II from CMB + BAO + \textit{HST} and CMB + BAO + \textit{HST} + GC, respectively.
             The orange continuous and dotted lines represent the Model I from CMB + BAO + \textit{HST}and CMB + BAO + \textit{HST} + GC, respectively.
             In drawing the graphs we have taken the best fit values from our analysis and Planck collaboration paper.}
    	\label{delta2}
\end{figure}

\section{Results}
\label{results}

Table \ref{tab2} summarizes the main results of the statistical analysis
carried out using two different combinations for the Model I, 
CMB + BAO + \textit{HST} and CMB + BAO +\textit{ HST} + GC. Fig.~\ref{Contours} shows the parametric space for some parameters 
of the Model I and its correlations. In both cases, we do not notice significant changes in the parameters $\xi_{\nu}$ and $c^2_{\rm eff}$ from standard prevision, i.e., $(\xi_{\nu}$, $c^2_{\rm eff}) = (0, \, 1/3)$. We note a small deviation on the viscosity parameter, $c^2_{\rm eff} \neq 1/3$, at 68 per cent CL in both analysis. Any value besides $c^2_{\rm eff} =1/3$ can be interpreted as an explicit coupling of the relativistic neutrino
(or some dark radiation) to a nonrelativistic particle species, e.g, cold dark matter \citep{R1,R2,R3,CNB5}.
In general terms, the presence of a dark radiation-dark matter interaction, the clustering properties of the dark radiation 
can be modified (see \cite{R1,R2,R3,CNB5} and references therein). That is, if dark radiation is composed of interacting particles,
the values of the parameters $c^2_{\rm eff}$ and $c^2_{\rm vis}$ can differ from the usual ones.
In this present work, we report that sound speed in the CNB rest frame is closed in the standard value, that is, 
$c^2_{\rm eff} = 1/3$ and $c^2_{\rm vis} \neq 1/3$ at 68 per cent CL, in both analysis.

In the standard scenario (three active neutrinos and considering effects related to
non-instantaneous neutrino decoupling), we have $N_{\rm eff} \simeq 3.046$. 
As previously introduced, the presence of a dark radiation is usually parametrized 
in the literature by $\Delta N_{\rm eff} \simeq N_{\rm eff} - 3.046$. 

From our results, we can note a small excess over $N_{\rm eff}$. More specifically,
we have $\Delta N_{\rm eff} \simeq 0.364 \, (0.614)$ from the best-fitting values for CMB + BAO + \textit{HST} (CMB + BAO +\textit{ HST} + GC). When evaluated the border at 95 per cent CL, we note $\Delta N_{\rm eff} < 0.794 \, (1.09)$ from CMB + BAO + \textit{HST} (CMB + BAO + \textit{HST} + GC. There are many candidates for dark radiation, for instance, sterile neutrinos \citep{R4}, thermal axions \citep{R5} and Goldstone bosons \citep{R6}. We can note that the constraints for $\Delta N_{\rm eff}$ is consistent with a partly thermalized sterile neutrino or a Goldstone boson from CMB + BAO +\textit{ HST}+ GC (best fit) and CMB + BAO + \textit{HST} (border 95 per cent CL).
A fully thermalized sterile neutrino is consistent at 95 per cent CL from CMB + BAO + \textit{HST} + GC.
About the neutrino mass scale, we have $\sum m_{\nu} < 0.36$ eV ($< 0.81$ eV) at 95 per cent CL from CMB + BAO + \textit{HST} (CMB + BAO +\textit{ HST} + GC). We see a variation around 0.45 eV when the GC data are added. 

Table \ref{tab3} summarizes the main results of the statistical analysis
carried out using two different combinations for the Model II, CMB + BAO + \textit{HST} and CMB + BAO + \textit{HST} + GC.
Fig.~\ref{Contours2} shows the parametric space for some parameters of the Model II.
In both analysis, we do not observe significant deviation of the degeneracy parameter $\xi_{\nu}$ from the null value.
However, in Model II, a small variation on $\Delta N_{\rm eff}$ can be noticed compared to the Model I.
We have $\Delta N_{\rm eff} \simeq 0.454 \, (0.604)$ from CMB + BAO + \textit{HST} (CMB + BAO + \textit{HST} + GC), 
respectively. A border at 95 per cent CL reads $\Delta N_{\rm eff} < 0.884 \, (< 1.17)$ for CMB + BAO + \textit{HST} (CMB + BAO + \textit{HST} + GC). Here, a partly thermalized sterile neutrino or a Goldstone boson can be accommodated in both analyzes.
Within the context of the Model II, when the GC data is added to CMB + BAO + \textit{HST}, we have a variation of around 0.34 eV 
on the neutrino mass scale.

The correlation between the extra relativistic degrees of freedom and the Hubble constant $H_0$ is well known.
Within the standard $\Lambda$CDM baseline, the \textit{Planck} collaboration \citep{Planck2015}
measured $H_0 = 67.27 \pm 0.66$ km s${}^{-1}$ Mpc${}^{-1}$, that is about
99 per cent CL deviations away from the locally measured value $H_0=73.24 \pm 1.74$ km s${}^{-1}$ Mpc${}^{-1}$, 
reported in \cite{riess}. The left panel of Fig.~\ref{HS1} showsl the likelihoods for $H_0$ resulting from the two cases analysed here. Changes in the central value of $H_0$ are not observed, and both cases return very similar fits
with $H_0 \simeq 70$ km s${}^{-1}$ Mpc${}^{-1}$. That intermediate value in comparison with the local and global constraints
can assuage the current tension on the Hubble constant.

In our analysis, we are take $\xi_{\nu}$ as a free parameter. In addition to interpreting $\Delta N_{\rm eff}$ 
only as a contribution due a some dark radiation, it is well known that the impact of the leptonic asymmetry increases the radiation energy density. Assuming three neutrino species with degenerated chemical potential $\xi_{\nu}$, we can write

\begin{eqnarray}
\label{xi1}
\Delta N_{\rm eff} = \Delta N^{\xi}_{\rm eff} + \Delta N^{\rm dr}_{\rm eff},  
\end{eqnarray}

where $\Delta N^{\xi}_{\rm eff}$, $\Delta N^{\rm dr}_{\rm eff}$ represents the contribution from the cosmological lepton asymmetry
and dark radiation, respectively. The increase via the leptonic asymmetry can be parametrized by

\begin{eqnarray}
\label{xi2}
\Delta N^{\xi}_{\rm eff} = \frac{90}{7} \Big( \frac{\xi_{\nu}}{\pi} \Big)^2 + \frac{45}{7} \Big( \frac{\xi_{\nu}}{\pi} \Big)^4.  
\end{eqnarray}

It is important to make clear that in all analyses, we are taking $N_{\rm eff}$ as free parameter, and we do not directly evaluating $\Delta N_{\rm eff}$ in our chains. Without loss of generality, we can evaluate the contribution in equation (\ref{xi2}) via the standard error propagation theory. We note, $\Delta N^{\xi}_{\rm eff} = 0.013 \pm 0.261 \, (0.0005 \pm 0.044)$ for the Model I from CMB + BAO + \textit{HST} (CMB + BAO + \textit{HST}+ GC). For the Model II, we have $\Delta N^{\xi}_{\rm eff} = 0.0032 \pm 0.127 \, ( 0.00052 \pm 0.046)$ 
for CMB + BAO + \textit{HST} (CMB + BAO + \textit{HST} + GC). Therefore, in general, from our analysis we can claim 
that the contribution from $\Delta N^{\xi}_{\rm eff}$ is very small, i.e., $\Delta N^{\xi}_{\rm eff} \ll \Delta N^{\rm dr}_{\rm eff}$ 
and $\Delta N_{\rm eff} \simeq \Delta N^{\rm dr}_{\rm eff}$.

Fig.~\ref{delta1}  (left panel) shows the linear neutrino perturbations as a function of the scale factor for three different scales. 
The solid and dashed lines represent models I, II, respectively, using the best fit from CMB + BAO + \textit{HST} + GC. 
Having an account of the physical variation of Model I ($c^2_{\rm eff}$ and $c^2_{\rm vis}$ from the best fit in table II) to 
Model II ($c^2_{\rm eff}=c^2_{\rm vis}=1/3$), this causes a very small change in the amplitude and phase of the density perturbations.
Variation in the degeneracy parameter does not significantly affect the perturbations. In the right panel (Fig.~\ref{delta2}),
we have a comparison of the effects on CMB TT of the extended models investigated here and the six parameter $\Lambda$CDM model from 
\textit{Planck} team \citep{Planck2015}. We can see that the theoretical prediction of the Model II up to $l \lesssim 3000$ is very similar to 
the six parameter $\Lambda$CDM model. The Model I shows variations around 7 per cent up to the range of the \textit{Planck} CMB TT data.  

\section{Conclusions}
\label{conclusions}

We have updated and improved the constraints on the neutrino properties within an 
extended $\Lambda$CDM $+ N_{\rm eff} + \sum m_{\nu} + c^2_{\rm  eff} + c^2_{\rm  vis} + \xi_{\nu}$ scenario 
using \textit{HST} and GC data as well as CMB measurements. We find that $c^2_{\rm  vis}$ can minimally deviate from its standard value 
at 68 per cent CL. A significant increase on $N_{\rm eff}$ can be seen, showing the possibility of presence of some dark radiation such as a partly thermalized sterile neutrino or a Goldstone boson when the GC data are added. 
The presence the GC data practically doubles the value of the neutrino mass scale (see table \ref{tab2}).
Cosmological constraints from GC can be affected from several systematics, such as the biased mass-observable relation, 
multiplicative shear bias, mean redshift bias, baryon feedback, and intrinsic alignment. Therefore, 
these small increases on $N_{\rm eff}$ that we noticed here could certainly due to systematic effects in GC data compilation. 
The constrains on $\sigma_8$ with and without GC are not really consistent, where we can easily see there a tension of 
approximately 2$\sigma$ CL on $\sigma_8$ (see Figures \ref{Contours} and \ref{Contours2}). For a discussion aof this (new physics or systematic effects), we refer to \cite{Kitching} and references therein. We do not find significant deviation for $c^2_{\rm  eff}$ and $\xi_{\nu}$. 
In the particular case of $\Lambda$CDM $+ N_{\rm eff} + \sum m_{\nu} + \xi_{\nu}$, 
no significant changes are observed in relation to the general case, and therefore the conclusions about the 
properties of the free parameters are the same. In both models, no mean deviations are found for the degeneracy parameter 
and $\xi_{\nu} \simeq 0$.

It is known that the neutrino properties can correlate in different ways with other
cosmological parameters. In the present work, we have considered the $\Lambda$CDM model 
to investigate the constraints on the properties of the relic neutrinos.  Recently, it has been discovered that the presence of massive neutrinos in cosmic dynamics can lead to small deviations of  the $\Lambda$CDM scenario \citep{C1,C2,C3,C4, C5}. Therefore, it is plausible to consider a parametric space extension by including neutrinos properties to models beyond the $\Lambda$CDM model (phenomenology of dark energy and 
modified gravity models). It can bring new perspectives in this direction.

\section*{Acknowledgments}

The authors thank the anonymous referee for his/her comments and suggestions.
Also, the authors are grateful to Suresh Kumar for the computational support with CLASS/Montepython code and
Alessandro Melchiorri for the critical reading of the manuscript.

\appendix

\section{Massive degenerate neutrinos}

At the very early Universe, neutrinos and antineutrinos of each flavour $\nu_i$ ($i = e,\mu,\tau$) behave
like relativistic particles and the energy density and pressure of one species of massive degenerate neutrinos and antineutrinos are 
described by (we use $\hbar = c = k_B$)

\begin{eqnarray}
 \rho_{\nu_i} +  \rho_{\bar{\nu_i}} = T^4_{\nu} \int \frac{d^3q}{2(\pi)^3} q^2 E_{\nu_i} (f_{\nu_i}(q) + f_{\bar{\nu_i}})) 
\end{eqnarray}

and 

\begin{eqnarray}
3 (p_{\nu_i} +  p_{\bar{\nu_i}}) = T^4_{\nu} \int \frac{d^3q}{2(\pi)^3} \frac{q^2}{E_{\nu_i}}(f_{\nu_i}(q) + f_{\bar{\nu_i}})),
\end{eqnarray}

where $E^2_{\nu_i} = q^2 + a^2 m_{\nu_i}$ is one flavour neutrino/antineutrinos energy and $q = a p$ is the comoving momentum.
The functions $f_{\nu_i}(q)$, $f_{\bar{\nu_i}}$ are the Fermi-Dirac phase space distributions given by

\begin{equation}
f_{\nu_i}(q) = \frac{1}{e^{E_{\nu_i}/T_{\nu} - \xi_{\nu}} + 1}, f_{\bar{\nu_i}}(q) = \frac{1}{e^{E_{\bar{\nu_i}}/T_{\nu} - \xi_{\bar{\nu}}} + 1}  
\end{equation}

where $\xi_{\nu} = \mu/T_{\nu}$ is the neutrino degeneracy parameter. The presence of a significant and non-null $\xi_{\nu}$ 
have some cosmological implication \citep{xi1,xi2,xi3,xi4,xi5,xi6,xi7,xi8,xi9,xi10,xi11,xi12,xi13,Dominik01,Dominik02}.


\begin{thebibliography}{}

\bibitem{CNB1} Betts S. et al., Development of a Relic Neutrino Detection Experiment at PTOLEMY: Princeton Tritium Observatory for Light, 
Early-Universe, Massive-Neutrino Yield (2013). [arXiv:1307.4738].

\bibitem{CNB2} Follin B., Knox L., Millea M., Pan, Z., 2015, Phys. Rev. Lett. 115, 091301. [arXiv:1503.07863]

\bibitem{Dolgov} Dolgov A.D., 2002, Phys. Rept., 370, 333. [arXiv:hep-ph/0202122]

\bibitem{Lesgourgues} Lesgourgues J., Pastor S., Phys. Rept., 429, 307. [arXiv:astro-ph/0603494]

\bibitem{Abazajian} Abazajian K.N. et al., 2015, Astropart. Phys., 63, 66. [arXiv:1309.5383]

\bibitem{Planck2015} Ade P.A.R. et al., 2016, A \& A, 594, A13, Planck collaboration. [arXiv:1502.01589]
 
\bibitem{book} Lesgourgues J., Mangano G., Miele G., Pastor S., 2013, Neutrino Cosmology, Cambridge University Press

\bibitem{CNB3} Archidiacono M., Calabrese E., Melchiorri A., 2011, Phys.Rev. D, 84, 123008. [arXiv:1109.2767]
%
 \bibitem{CNB4} Archidiacono M., Giusarma E., Melchiorri A., Mena O., 2012, Phys.Rev. D, 86, 043509. [arXiv:1206.0109]
%
\bibitem{CNB5} Diamanti R., Giusarma E., Mena O., Archidiacono M., Melchiorri A., 2013, Phys.Rev. D, 87, 063509. [arXiv:1212.6007]

\bibitem{CNB6}  Archidiacono M.,  Giusarma, E.,  Melchiorri A., and  Mena O., 2013, Phys.Rev. D, 87, 103519. [arXiv:1303.0143]
%
\bibitem{CNB7} Gerbino M., Valentino E.D., Said, N., 2013, Phys.Rev. D, 88, 063538. [arXiv:1304.7400]

\bibitem{CNB8} Trotta R., Melchiorri A., 2005, Phys. Rev. Lett., 95, 011305. [arXiv:astro-ph/0412066]

\bibitem{CNB9} Krauss L.M., Long A.J., 2016, JCAP, 07, 002. [arXiv:1604.00886]

\bibitem{CNB10} Sellentin E., Durrer R., 2015, Phys. Rev. D, 92, 063012. [arXiv:1412.6427] 

\bibitem{Benjamin} Audren B. et al., 2015, JCAP, 03, 036. [arXiv:1412.5948]

\bibitem{Hu1} Hu W., 1998, Astrophys. J., 506, 485. [arXiv:astro-ph/9801234] 
%
\bibitem{Hu2} Hu W.,  Eisenstein D.J.,  Tegmark M.,  White M.J., 1999 Phys. Rev. D, 59, 023512. [arXiv:astro-ph/9806362]
%
\bibitem{xi1}  Serpico P.D., Raffelt G.G., 2005, Phys. Rev. D, 71, 127301. [astro-ph/0506162]

\bibitem{xi2} Simha V., Steigman G., 2008, JCAP, 08, 011. [arXiv:0806.0179]

\bibitem{xi3} Popa L.A., Vasile A., 2008, JCAP, 06, 028. [arXiv:0804.2971]
 
\bibitem{Dominik01}  D. J. Schwarz and M. Stuke, 2013, New J. Phys. 15 033021. [arXiv:1211.6721]

\bibitem{xi4} Wong Y.Y., 2002, Phys. Rev. D, 66, 025015. [hep-ph/0203180]

\bibitem{xi5} Abazajian K.N.,  Beacom J.F., Bell N.F., 2002,  Phys. Rev. D,  66, 013008. [astro-ph/0203442]

\bibitem{xi6} Hu W., Scott D., Sugiyama N., White M., 1995, Phys. Rev. D,  52, 5498. [astro-ph/9505043]
%
\bibitem{xi7} Lesgourgues J., Pastor S., 1999, Phys. Rev. D, 60, 103521. [hep-ph/9904411]

\bibitem{xi8} Lattanzi M., Ruffni R., Vereshchagin G.V., 2005, Phys. Rev. D,  72, 063003. [astro-ph/0509079]

\bibitem{xi9} Ichiki K., Yamaguchi M., Yokoyama J., 2007, Phys. Rev D,  75, 084017. [hep-ph/0611121]

\bibitem{xi10} Kinney W.H., Riotto A., 1999, Phys. Rev. Lett.,  83, 3366. [hep-ph/9903459] 

\bibitem{xi11} Caramete A., Popa L.A., 2014, JCAP,  02, 012. [arXiv:1311.3856]

\bibitem{xi12} Mangano. G, Miele. G, Pastor. S, Pisanti.O, Sarikas. S, 2012, Phys. Lett. B 708 1. [arXiv:1110.4335]

\bibitem{xi13} Castorina. E et al., 2012, Phys. Rev. D 86 023517. [arXiv:1204.2510]


\bibitem{Dominik02}  Oldengott I. M and Schwarz D. J, 2017, Europhys. Lett., 119, 29001. [arXiv:1706.01705].
%
 \bibitem{Ma_Bertschinger}  Ma C.P.,  Bertschinger E., 1995, Astrophys. J., 7, 455. [astro-ph/9506072]

\bibitem{Lewis}  Lewis A., Challinor A., 2002, Phys.Rev. D, 66, 023531. [arXiv:astro-ph/0203507]
%
\bibitem{Komatsu} Shoji M., Komatsu E., 2010, Phys.Rev. D, 81, 123516. [arXiv:1003.0942] 

\bibitem{Lesgourgues2} Lesgourgues J., Tram T., 2011, JCAP,  1109, 032. [arXiv:1104.2935] 
 
 \bibitem{bao1} Beutler F. et  al., 2011, MNRAS, 416, 3017. [arXiv:1106.3366] 
 
 \bibitem{bao2}  Ross A.J., Samushia L., Howlett C., Percival W.J., Burden A., Manera M., 2015, MNRAS, 449, 835. [arXiv:1409.3242]
 
\bibitem{bao3} Anderson L. et  al., 2014, MNRAS, 441, 24. [arXiv:1312.4877] 
 
\bibitem{bao4} Font-Ribera A. et  al., 2014, JCAP, 5, 27. [arXiv:1311.1767] 

\bibitem{bao5} Nunes R.C., Pan S., Saridakis E.N., 2016, Phys. Rev. D, 94, 023508. [arXiv:1503.04113]  
 
\bibitem{riess} Riess A.G. et al., 2016, Astrophys. J., 826, 56. [arXiv:1604.01424]

\bibitem{Bernal} Bernal J. L, Verde. L, Cuesta A. J., 2016, JCAP 02 059. [arXiv:1511.03049] 
 
\bibitem{Ruiz} Ruiz E. J and D. Huterer, 2015, Phys. Rev. D 91 6,063009. [arXiv:1410.5832]
 
\bibitem{CG1} Tinker J.L. et al., 2012, Astrophys. J.,  745, 16. [arXiv:1104.1635]

\bibitem{CG2} Rozo E. et al.,  2010, Astrophys. J., 708, 645. [arXiv:0902.3702]

\bibitem{CG3} Henry J.P. et al.,  2009, Astrophys. J.,  691, 1307. [arXiv:0809.3832]

\bibitem{CG4} Benson B.A. et al., 2013, Astrophys. J.,  763, 147. [arXiv:1112.5435] 

\bibitem{CG5} Hajian A. et al., 2013, JCAP, 11, 64. [arXiv:1309.3282]
%
\bibitem{CG6}  Ade et al. P.A.R., 2010, MNRAS, 406, 1759. [arXiv:0909.3098]
%
\bibitem{CG7} Ade et al. P.A.R, 2014, A\&A 571 A20. [arXiv:1303.5080]

\bibitem{CG8} Vikhlinin A. et al., 2009,  Astrophys. J.,  692, 1060. [arXiv:0812.2720].

\bibitem{CG9} Heymans C. et al., 2013, MNRAS, 432, 2433. [arXiv:1303.1808] 

\bibitem{class} Blas D., Lesgourgues J., Tram T., 2011, JCAP, 07, 034. [arXiv:1104.2933]  

\bibitem{monte} Audren B., Lesgourgues J., Benabed K., Prunet S., 2013, JCAP, 02, 001. [arXiv:1210.7183] 
 
\bibitem{Gelman} Gelman A., Rubin D., 1992, Inference from iterative simulation using multiple sequences, Statistical Science, 7, 457 

\bibitem{R1} Oldengott I. M., Rampf C., Wong Y. Y. Y., 2015, JCAP, 1504, 04, 016. [arXiv:1409.1577] 

\bibitem{R2} Smith T.L., Das S., Zahn O., 2012, Phys. Rev. D,  85, 023001. [arXiv:1105.3246] 

\bibitem{R3} Wilkinson R.J., Boehm C., Lesgourgues J., 2014, JCAP, 05, 011. [arXiv:1401.7597] 

\bibitem{R4} Archidiacono M. et al., 2016, JCAP, 08, 067. [arXiv:1606.07673]

\bibitem{R5} Valentino E.D., Melchiorri A., Mena, O., 2013, JCAP, 1311, 018. [arXiv:1304.5981] 

\bibitem{R6} Weinberg S., 2013, Phys. Rev. Lett., 110, 241301. [arXiv:1305.1971] 
 
\bibitem{Kitching}  Kitching T. D, Verde L., Heavens A. F, Jimenez R., Mon. Not. Roy. Astron. Soc., 2016, 
 459 1. [arXiv:1602.02960]

\bibitem{C1} Valentino E.D., Melchiorri A., Silk J., 2016, Phys. Lett. B, 761. [arXiv:1606.00634] 

\bibitem{C2} Valentino E.D., Melchiorri A., Linder E.V., Silk J., 2017, Phys. Rev. D, 96, 023523. [arXiv:1704.00762]

\bibitem{C3} Kumar S., Nunes R.C.,  2016, Phys. Rev. D, 94, 123511. [arXiv:1608.02454] 
 
\bibitem{C4} Kumar S., Nunes R.C., 2017, to appear in Phys. Rev. D. [arXiv:1702.02143]  
 
\bibitem{C5} Yang W., Nunes R. C, Pan S., Mota D. M., 2017, Phys. Rev. D 95, 103522. [arXiv:1703.02556]. 

 
 

 
 
\end{thebibliography}
\end{document}